\documentclass[preprint,12pt]{elsarticle}

\usepackage{amssymb}
\usepackage{inputenc}
\usepackage{algorithmic}
\usepackage{algorithm}

\usepackage[caption=false]{subfig}
\usepackage{geometry}
\begin{document}

\begin{frontmatter}

%% Title, authors and addresses

%% use the tnoteref command within \title for footnotes;
%% use the tnotetext command for the associated footnote;
%% use the fnref command within \author or \address for footnotes;
%% use the fntext command for the associated footnote;
%% use the corref command within \author for corresponding author footnotes;
%% use the cortext command for the associated footnote;
%% use the ead command for the email address,
%% and the form \ead[url] for the home page:
%%
%% \title{Title\tnoteref{label1}}
%% \tnotetext[label1]{}
%% \author{Name\corref{cor1}\fnref{label2}}
%% \ead{email address}
%% \ead[url]{home page}
%% \fntext[label2]{}
%% \cortext[cor1]{}
%% \address{Address\fnref{label3}}
%% \fntext[label3]{}

\title{Merge Path - A Visually Intuitive Approach to Parallel Merging}

%% use optional labels to link authors explicitly to addresses:
 \author[*,@,!]{Oded Green}
 \author[@]{Saher Odeh}
 \author[@]{Yitzhak Birk}
 \address[*]{College of Computing, Georgia Institute of Technology\\ Atlanta, Georgia, USA}
 \address[@]{Electrical Engineering, Technion - Israel Institute of Technology \\ Haifa, Israel}
	\address[!]{Corresponding author : ogreen@gatech.edu}

\begin{abstract}
Merging two sorted arrays is a prominent building block for sorting and other functions. Its efficient parallelization requires balancing the load among compute cores, minimizing the extra work brought about by parallelization, and minimizing inter-thread synchronization requirements. Efficient use of memory is also important. 

We present a novel, visually intuitive approach to partitioning two input sorted arrays into pairs of contiguous sequences of elements, one from each array, such that 1) each pair comprises any desired total number of elements, and 2) the elements of each pair form a contiguous sequence in the output merged sorted array. While the resulting partition and the computational complexity are similar to those of certain previous algorithms, our approach is different, extremely intuitive, and offers interesting insights. Based on this, we present a synchronization-free, cache-efficient merging (and sorting) algorithm. 

While we use a shared memory architecture as the basis, our algorithm is easily adaptable to additional architectures. In fact, our approach is even relevant to cache-efficient sequential sorting. The algorithms are presented, along with important cache-related insights.

\end{abstract}

\begin{keyword}
%% keywords here, in the form: keyword \sep keyword

Parallel algorithms \sep Parallel processing \sep Merging \sep Sorting 

%% MSC codes here, in the form: \MSC code \sep code
%% or \MSC[2008] code \sep code (2000 is the default)

\end{keyword}

\end{frontmatter}

%%
%% Start line numbering here if you want
%%
% \linenumbers

%% main text

\section{Introduction}

Merging two sorted arrays, $A$ and $B,$ to form a sorted output array $S$ is an important utility, and is the core the of merge-sort algorithm \cite{Cormen2001} . Additional uses include joining the results of database queries and merging adjacency lists of vertices in graph contractions.

The merging (e.g., in ascending order) is carried out by repeatedly comparing the smallest (lowest-index) as-yet unused elements of the two arrays, and appending the smaller of those to the result array. 

Given an (unsorted) \textit{N}-element array, merge-sort comprises a sequence of ${{\log }_2 N\ }$ merge rounds: in the first round, \textit{N/2} disjoint pairs of adjacent elements are sorted, forming \textit{N/2} sorted arrays of size two. In the next round, each of the \textit{N/4} disjoint pairs of two-element arrays is merged to form a sorted 4-element array. In each subsequent round, array pairs are similarly merged, eventually yielding a single sorted array. 

Consider the parallelization of merge-sort using $p$ compute cores (or processors or threads, terms that will be used synonymously). Whenever $N\gg p$, the early rounds are trivially parallelizable, with each core assigned a subset of the array pairs. This, however, is no longer the case in later rounds, as only few such pairs remain. Consequently and since the total amount of computation is the same in all rounds, effective parallelization requires the ability to parallelize the merging of two sorted arrays

An efficient Parallel Merge algorithm must have several salient features, some of which are required due to the very low compute to memory-access ratio: 1) equal amounts of work for all cores; 2) minimal inter-core communication (platform-dependent ramifications); 3) minimum excess work (for parallelizing, as well as replication of effort); and 4) efficient access to memory (high cache hit rate and minimal cache-coherence overhead). Coherence issues may arise due to concurrent access to the same address, but also due to concurrent access to different addresses in the same cache line (false sharing). Memory issues have platform-dependent manifestations.

A na{\"i}ve approach to parallel merge would entail partitioning each of the two arrays into equal-length contiguous sub-arrays and assigning a pair of same-size sub arrays to each core. Each core would then merge its pair to form a single sorted array, and those would be concatenated to yield the final result. Unfortunately, this is incorrect. (To see this, consider the case wherein all the elements of \textit{A} are greater than all those of \textit{B.}) So, correct partitioning is the key to success.

In this paper, we present a parallel merge algorithm for Parallel Random Access Machines (PRAM), namely shared-memory architectures that permit concurrent (parallel) access to memory. PRAM systems are further categorized as CRCW, CREW, ERCW or EREW, where C, E, R and W denote concurrent, exclusive, read and write, respectively. Our algorithm assumes CREW, but can be adapted to other variants. Also, complexity calculations assume equal access time of any core to any address, but this is not a requirement.

Our algorithm is load-balanced, lock-free, requires a negligible amount of excess work, and is extended to a memory-efficient version. Being lock-free, the algorithm does not rely on a set of atomic instructions of any particular platform and therefore can be easily applied. The efficiency of memory access is also not confined to one kind of architecture; in fact, the memory access is efficient for both private- and shared-cache architectures. 

We show a correspondence between the merge operation and the traversal of a path on a grid, from the upper left corner to the bottom right corner and going only rightward or downward. This greatly facilitates the comprehension of parallel merge algorithms. By using this path, dubbed \textit{Merge Path}, one can divide the work equally among the cores. Most important, we parallelize the partitioning of the merge path.

Our actual basic algorithm is similar to that of \cite{Deo1991} , but is more intuitive and conceptually simpler. Furthermore, using insights from the aforementioned geometric correspondence, we develop a new cache-efficient merge algorithm, and use it for a memory-efficient parallel sort algorithm. 

The remainder of the paper is organized as follows. In Section II, we present the Merge Path, the Merge Matrix and the relationship between them. These are used in Section III to develop parallel merge and sort algorithms. Section IV introduces cache-related issues and presents a cache-efficient merge algorithm. Section V discusses related work and Section VI presents experimental results of our two new parallel algorithms on two systems.Section VII offers concluding remarks.

\section{ Merge Path}
\begin{figure*}%[t]%
\centering
  \subfloat[]{\includegraphics[width=0.5\columnwidth]{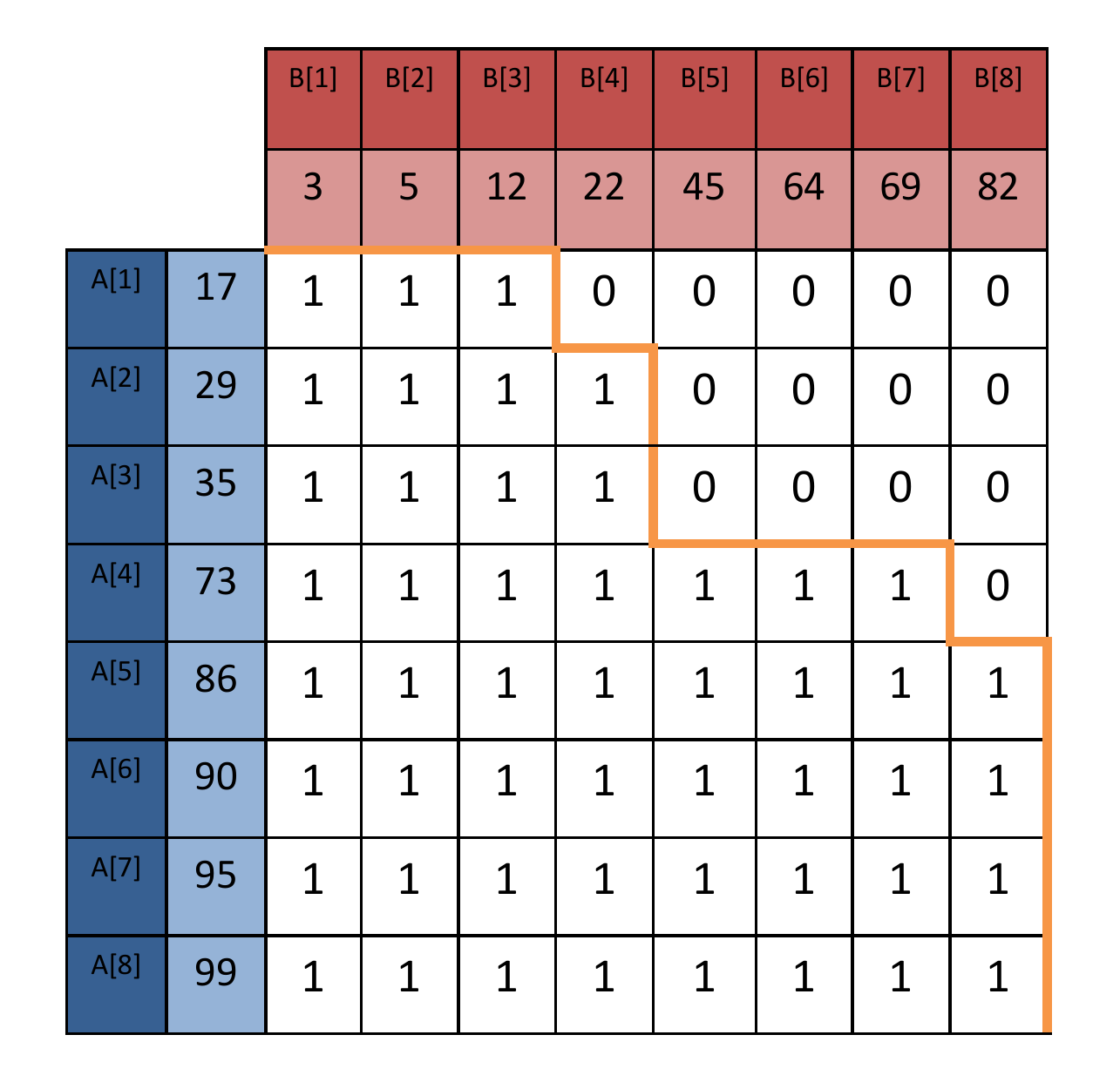}}
  \subfloat[]{\includegraphics[width=0.5\columnwidth]{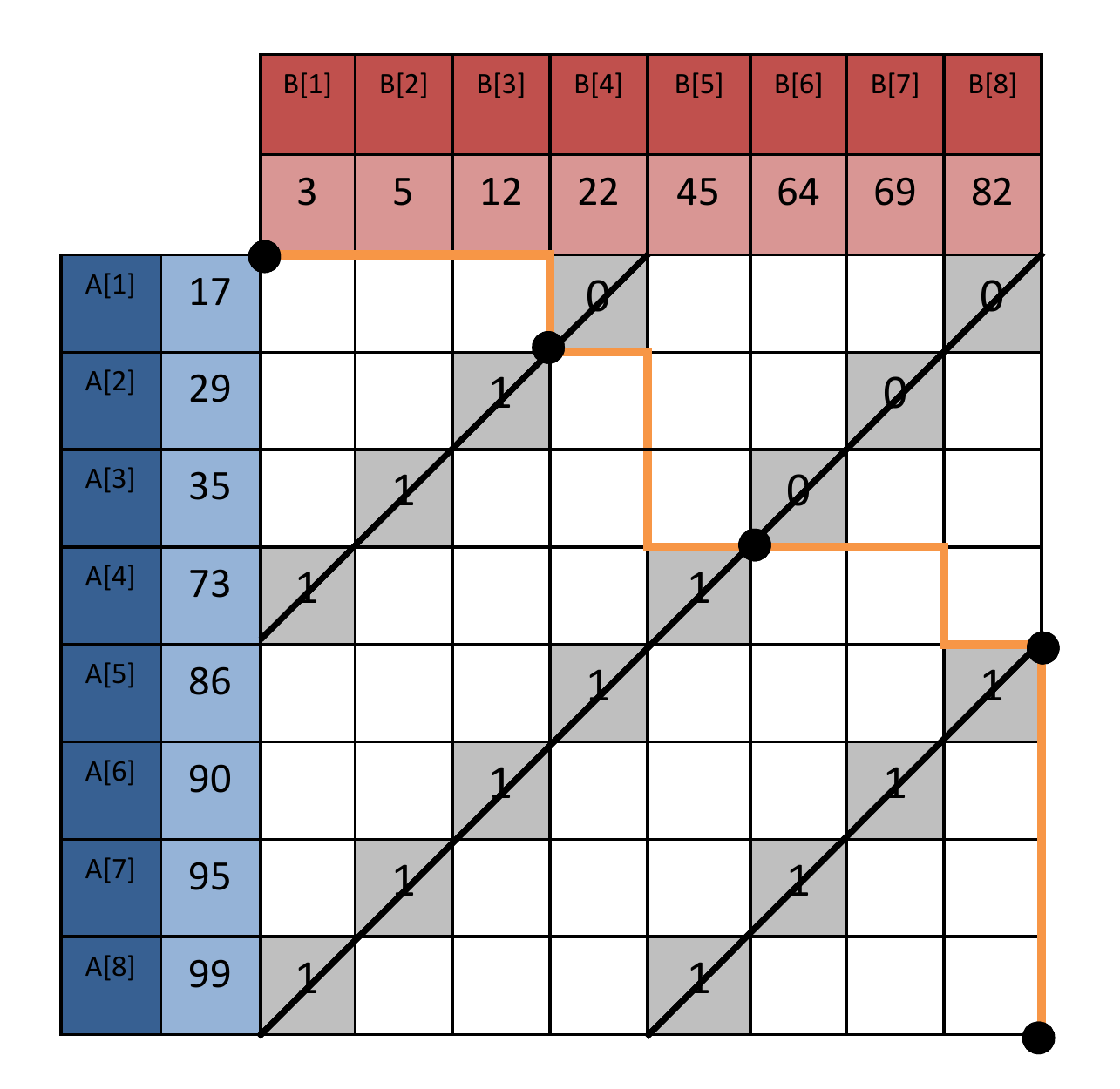}}
  \label{fig:MovingWindow}
  \caption{Merge Matrix and Merge Path. (a) The Merge Matrix is shown with all the values explicitly computed. The Merge Path is on the boundary between the zeros and the ones. (b) The cross diagonals in a Merge Matrix are used to find the points of change between the ones and the zeros, i.e., the intersections with the Merge Path.} 
\end{figure*}

\subsection{ Construction and basic properties}

Consider two sorted arrays, \textit{A} and \textit{B}, with \textit{\textbar A\textbar } and \textit{\textbar B\textbar } elements, respectively. (The lengths of the arrays may differ: \textit{\textbar A\textbar } $\neq$ \textit{\textbar B\textbar }  ) Without loss of generality, assume that they are sorted in ascending order. As depicted in Fig. 1 (a) (ignore the contents of the matrix), create a column comprising \textit{A}'s elements and a Row comprising \textit{B}'s elements, and an $|A|x|B|$ matrix \textit{M}, each of whose rows (columns) corresponds to an element of \textit{A} (\textit{B}). We refer to this matrix as the \textit{Merge Matrix}. A formal definition and additional details pertaining to \textit{M} will be provided later.

Next, let us merge the two arrays: in each step, pick the smallest (regardless of array) yet-unused array element. Alongside this process, we construct the \textit{Merge Path}. Referring again to Fig. 1 (a), start in the upper left corner of the grid, i.e., at the upper left corner of \textit{M[1,1]}. If \textit{A[1]$>$B[1]}, move one position to the right\textit{; }else move one position downward\textit{. }Continue similarly as follows: consider matrix position \textit{(i,j) }whose upper left corner is the current end of the merge path: if  \textit{A[i]$>$B[j]}, move one position to the right and increase $j$\textit{; }else move one position downward and increase $i$; having reached the right or bottom edge of the grid, proceed in the only possible direction. Repeat until reaching the bottom right corner. 

The following four lemmas follow directly from the construction of the Merge Path:

\textbf{Lemma 1: }Traversing a Merge Path from beginning to end, picking in each rightward step the smallest yet-unused element of \textit{B,} and in each downward step the smallest yet-unused element of \textit{A}, yields the desired merger.

\hfill $\blacksquare$

\textbf{Lemma 2:} Any contiguous segment of a Merge Path is composed of a contiguous sequence of elements of \textit{A }and of a contiguous sequence of elements of \textit{B}.     \hfill      $\blacksquare$

\textbf{Lemma 3:} Non-overlapping segments of a merge path are composed of disjoint sets of elements, and vice versa.                \hfill $\blacksquare$

\textbf{Lemma 4: }Given two non-overlapping segments of a merge path, all array elements composing the later segment are greater than or equal to all those in the earlier segment.                         \hfill $\blacksquare$

\textbf{Theorem 5:} Consider a set of element-wise disjoint sub-array pairs (one, possibly empty sub-array of \textit{A} and one, possibly empty sub-array of \textit{B}), such that each such pair comprises all elements that, once sorted,  jointly form a contiguous segment of a merge path. It is claimed that these array pairs may be merged in parallel and the resulting merged sub-arrays may be concatenated according to their order in the merge path to form a single sorted array.

\textit{Proof:} By Lemma 1, the merger of each sub-array pair forms a sorted sub-array comprising all the elements in the pair. From Lemma 2 it follows that each such sub-array is composed of elements that form a contiguous sub-array of their respective original arrays, and by Lemma 3 the given array pairs correspond to non-overlapping segments of a merge path. Finally, by Lemma 4 and the construction order, all elements of a higher-indexed array pair are greater than or equal to any element of a lower-indexed one, so concatenating the merger results yields a sorted array.            \hfill $\blacksquare$

\textbf{Corollary 6:} Any partitioning of a given Merge Path of input arrays \textit{A} and \textit{B} into non-overlapping segments that jointly comprise the entire path, followed by the independent merger of each corresponding sub-array pair and the concatenation of the results in the order of the corresponding Merge-Path segment produces a single sorted array comprising all the elements of \textit{A} and \textit{B}.                    \hfill $\blacksquare$

\textbf{Corollary 7: }Partitioning a Merge Path into equisized segments and merging the corresponding array pairs in parallel balances the load among the merging processors. 

\textit{Proof: }each step of a Merge Path requires the same operations (read, compare and write), regardless of the outcome.                     \hfill $\blacksquare$

Equipped with the above insights, we next set out to find an efficient method for partitioning the Merge Path into equal segments. The challenge, of course, is to do so without actually constructing the Merge Path, as its construction is equivalent to carrying out the entire merger. Once again using the geometric insights provided by Fig. 1 (b), we begin by exposing an interesting relationship between positions on any Merge Path and cross diagonals (ones slanted upward and to the right) of the Merge Matrix \textit{M}. Next, we define the contents of a Merge Matrix and expose an interesting property involving those. With these two building blocks at hand, we construct a simple method for parallel partitioning of any given Merge Path into equisized segments. This, in turn, enables parallel merger.

\subsection{ The Merge Path and cross diagonals}

\textbf{Lemma 8: }Regardless of the route followed by a Merge Path, and thus regardless of the contents of \textit{A} and \textit{B}, the \textit{i'}th point along a Merge Path lies on the \textit{i}'th cross diagonal of the grid and thus of the Merge Matrix \textit{M}.

\textit{Proof:} each step along the Merge Path is either to the right or downward. In either case, this results in moving to the next cross diagonal.              \hfill $\blacksquare$

\textbf{Theorem 9:} Partitioning a given merge path into \textit{p }equisized contiguous segments is equivalent to finding its intersection points with \textit{p-1} equispaced cross diagonals of \textit{M}, 

\textit{Proof:} follows directly from Lemma 8.                     \hfill $\blacksquare$

\subsection{ The Merge Matrix -- content \& properties}

\textbf{Definition 1:} A binary merge matrix $M$ of $A,B$ is a Boolean two dimensional matrix of size $\left|A\right|\times \left|B\right|$ such that$:$ 
\[M\left[i,j\right]=\left\{ \begin{array}{c}
 \begin{array}{cc}
1 & A\left[i\right]>B\left[j\right] \end{array}
 \\ 
 \begin{array}{cc}
0 & otherwise \end{array}
 \end{array}
\right..\]

\textbf{Proposition 10:} Let $M$ be a binary merge matrix. Then,  $M\left[i,j\right]=1$$\Rightarrow$$\forall k,m:i\le k\le \left|A\right|\wedge 1\le m\le j,\ M\left[k,m\right]=1$

\textit{Proof:} If $M\left[i,j\right]=1$ then according to definition 1, $A\left[i\right]>B[j]$. $k\ge i\Rightarrow A\left[k\right]\ge A\left[i\right]$ ($A$ is sorted). $j\ge m\Rightarrow B\left[j\right]\ge B\left[m\right]$ ($B$ is sorted). $A\left[k\right]\ge A\left[i\right]>B\left[j\right]\ge B\left[m\right]$ and according to definition 1,  $M\left[k,m\right]=1$.                       \hfill $\blacksquare$

\textbf{Proposition 11:} Let $M$ be a binary merge matrix. If $M\left[i,j\right]=0$, then  $\forall k,m\ s.t.\left({\rm 1}\le k<i\right)\wedge (j\le m\le \left|B\right|)$, $M\left[k,m\right]=0$.

\textit{Proof:} Similar to the proof of proposition 10.                    \hfill $\blacksquare$

\textbf{Corollary 12. }The entries along any cross diagonal of \textit{M} form a monotonically non-increasing sequence.                       \hfill $\blacksquare$

\subsection{ The Merge Path and the Merge Matrix}

Having established interesting properties of both the Merge Path and the Merge Matrix, we now relate the two, and use \textit{P(M)} to denote the Merge Path corresponding to Merge Matrix \textit{M}.

\textbf{Proposition 13:} Let $\left(i,j\right)$ be the highest point on a given cross diagonal $M$ such that $M\left[i,j-1\right]=1$ if exists, otherwise let $\left(i,j\right)$ be the lowest point on that cross diagonal. Then, $P\left(M\right)\ $passes through $\left(i,j\right)$. This is depicted in Fig 2.

\textit{Proof:} by induction on the points on the path.

\textit{Base:} The path starts at $\left(1,1\right)$. The cross diagonal that passes through $\left(1,1\right)$ consists only of this point; therefore, it is also the lowest point on the cross diagonal.\textbf{}

\textit{Induction step:} assume the correctness of the claim for all the points on the path up to point $\left(i,j\right)$. Consider the next point on $P\left(M\right)$. Since the only permissible moves are $Right$ and $Down$, the next point can be either $\left(i,j+1\right)$ or $\left(i+1,j\right),$ respectively. 

Case 1: $Right$ move. The next point is $\left(i,j+1\right)$. According to Definition 1, $M\left[i,j\right]=1$. According to the induction assumption, either $i=1$ or $M\left[i-1,j\right]=0$. If $i=1$ then the new point is the highest point on the new cross diagonal such that $M\left[i,j\right]=1$. Otherwise, $M\left[i-1,j\right]=0$. According to Proposition 11, $M\left[i-1,j+1\right]=0$. Therefore, $\left(i,j+1\right)$ is the highest point on its cross diagonal at which $M\left[i,j\right]=1$.

\begin{algorithm}[t]
  \caption{$Parallel Merge(A,B,S,p)$}
  \begin{algorithmic}
    \FOR{$i=1$ $to$ $p$ $\mathbf{parallel}$}
    	\STATE $DiagonalNum \leftarrow (i-1) \cdot (|A|+|B|)/p +1$
    	\STATE $length \leftarrow (|A|+|B|)/p$
			\STATE $a_{i,start},b_{i,start} \leftarrow $ Diagonal Intersection (A,B,i,p) // Algorithm \ref{alg:diagInter}
    	\STATE $s_{i,start} \leftarrow (i-1) \cdot (|A|+|B|)/p +1$
			\STATE $Merge(A,a_{i,start},B,b_{i,start},S,s_{i,start},length)$
    \ENDFOR
		\STATE $\mathbf{Barrier}$
		\label{alg:MergePath}
  \end{algorithmic}			
\end{algorithm}

\begin{algorithm}[t]

  \caption{$Diagonal Intersection(A,B,thread_{id},p)$ -  Algorithm for finding intersection of the Merge Path and the cross diagonals.}
		\begin{algorithmic}
%    \STATE $A_{diag}[{threads}] \Leftarrow |A|$
%    \STATE $B_{diag}[{threads}] \Leftarrow |B|$

%    \FOR{$\mathbf{each}$ $i$ $\mathbf{in}$ $threads$ $\mathbf{in}$ $\mathbf{parallel}$}
      \STATE $diag \Leftarrow i \cdot (|A| + |B|)$ / $p$
      \STATE $a_{top} \Leftarrow diag > |A|$ ? $|A|$ : $diag$
      \STATE $b_{top} \Leftarrow diag > |A|$ ? $diag - |A|$ : 0
      \STATE $a_{bottom} \Leftarrow b_{top}$

      \WHILE{true}
			\STATE $offset \Leftarrow (a_{top} - a_{bottom}) / 2$
			\STATE $a_i \Leftarrow a_{top} - offset$
			\STATE $b_i \Leftarrow b_{top} + offset$
			\IF{$A[a_i] > B[b_i-1]$}
				\IF{$A[a_i-1] \leq B[b_i]$}
					\STATE $a_{start} \Leftarrow a_i$
					\STATE $b_{start}  \Leftarrow b_i$
					\STATE $\mathbf{break}$
				\ELSE
					\STATE $a_{top} \Leftarrow a_i - 1$
					\STATE $b_{top} \Leftarrow b_i + 1$
				\ENDIF
			\ELSE
				\STATE $a_{bottom} \Leftarrow a_i + 1$
			\ENDIF
					\ENDWHILE
%    \ENDFOR
		\RETURN $\{a_{start},b_{start} \}$
  \end{algorithmic}
%  \caption{Pseudo code for parallel Merge Path algorithm with an emphasis on the partitioning stage.}
\label{alg:diagInter}
		
\end{algorithm}

Case 2: the move was $Down$, so the next point is $\left(i+1,j\right)$. According to Definition 1, $M\left[i,j\right]=0$. According to the induction assumption, either $j=1$ or $M\left[i,j-1\right]=$1. If $j=1$ then the new point is the lowest point in the new cross diagonal. Since $M\left[i,j\right]=0$ and according to Proposition 11, the entire cross diagonal is $0$. Otherwise, $M\left[i,j-1\right]=1$. According to Proposition 10, $M\left[i+1,j-1\right]=1$. Therefore, $\left(i,j+1\right)$ is the highest point on its cross diagonal at which $M\left[i+1,j-1\right]=1$.                         \hfill $\blacksquare$

\textbf{Theorem 14: }Given sorted input arrays \textit{A} and \textit{B}, they can be partitioned into \textit{p} pairs of sub-arrays corresponding to \textit{p} equisized segments of the corresponding merge path. The \textit{p-1} required partition points can be computed independently of one another (optionally in parallel), in at most log${}_{2}$(min(\textbar \textit{A\textbar ,\textbar B\textbar })) steps per partition point, with neither the matrix nor the path having to actually be constructed.

\textit{Proof:} According to Theorem 9, the required partition points are the intersection points of the Merge Path with \textit{p-1} equispaced (and thus content-independent) cross diagonals of \textit{M}. According to Corollary 12 and Proposition 13, each such intersection point is the (only) transition point between `1's and `0's along the corresponding cross diagonal. (If the cross diagonal has only `0's or only `1's, this is the uppermost and the lowermost point on it, respectively.) Finding this partition point, which is the intersection of the path and the cross diagonal, can be done by way of a binary search on the cross diagonal, whereby in each step a single element of \textit{A} is compared with a single element of \textit{B}. Since the length of a cross diagonal is at most min(\textbar \textit{A\textbar ,\textbar B\textbar }), at most log${}_{2}$(min(\textbar \textit{A\textbar ,\textbar B\textbar })) steps are required. Finally, it is obvious from the above description that neither the Merge Path nor the Merge Matrix needs to be constructed and that the \textit{p-1} intersection points can be computed independently and thus in parallel.                   \hfill $\blacksquare$

\section{Parallel Merge and Sort}

Given two input arrays \textit{A} and \textit{B} parallel merger is carried out by \textit{p} processors according to Algorithm 1. For the sake of brevity we do not present the pseudo code for a sequential merge.

\textbf{Remark. }Note that no communication is required among the cores: they write to disjoint sets of addresses and, with the exception of reading in the process of finding the intersections between the Merge Path and the diagonals, read from disjoint addresses. Whenever \textit{\textbar A\textbar +\textbar B\textbar $>$$>$p}, which is the common case, this means that concurrent reads from the same address are rare.\textbf{ }

Summarizing the above, the time complexity of the algorithm for $|A|+|B|=N$ and \textit{p} processors is $O\left(N/p+{\log  \left(N\right)\ }\right),\ $and the work complexity is $O(N+p\cdot {\log  N\ }$). For $p<N/log(N),\ \ $this algorithm is considered to be optimal. In the next section, we address the issue of efficient memory (cache) utilization. 

Finally, merge-sort can be employ Parallel Merge to carry out each of ${{\log }_2 N\ }$ rounds. The rounds are carried out one after the other.

The time complexity of this Parallel Merge-Sort is: 
$O(N/p \cdot log(N/p) + N/p \cdot log(p) + log(p) \cdot log  (N) = 
O(N//p \cdot log(N) + log(p) \cdot log(n)$

In the first expression, the first component corresponds to the sequential sort carried out concurrently by each core on \textit{N/p} input elements, and the two remaining ones correspond to the subsequent rounds of parallel merges.

\section{ Cache Efficient Merge Path}
In the remainder of this section, we examine the cache efficiency issue in conjunction with our algorithm, offering important insights, exploring trade-offs and presenting our approaches. Before continuing along this path, however, let us digress briefly to discuss relevant salient properties of hierarchical memory in general, and particularly in shared-memory environments. We have kept this discussion short and only present relevant properties of the cache. For additional information on caches we refer the reader to Stenstrom \cite{stenstrom1990survey} for a brief survey on cache coherence, Conway \emph{et al.} \cite{conway2010cache} for cache design considerations for modern caches, and Lam \emph{et al.} \cite{lam1991cache} for performance of blocked algorithms that take the cache into consideration.

\subsection{ Overview}

The rate at which merging and sorting can be performed even in memory (as opposed to disk), is often dictated by the performance of the memory system rather than by processing power. This is due to the fact that these operations require a very small amount of computing per unit of data, and the fact that only a small amount of memory, the cache, is reasonably fast. (The next level in the memory hierarchy typically features a ten-fold higher access latency as well as coarser memory-management granularity.) Parallel implementation on a shared memory system further aggravates the situation for several reasons: 1) the increased compute power is seldom matched by a commensurate increase in memory bandwidth, at least beyond the 1${}^{st}$-level or 2${}^{nd}$-level cache, 2) the cores potentially share a 2${}^{nd}$ or 3${}^{rd}$ level cache, and 3) cache coherence mechanisms can present an extremely high overhead. In this section, we address the memory issues.

 Assuming large arrays (relative to cache size) and merge-sort, it is clear that data will have to be brought in multiple times (${{\log }_2 N\ }$ times, one for each level of the merge tree), so we again focus on merging a pair of sorted arrays. 

%The reader may skip this and proceed directly to 4.3.

\noindent 

\subsection{ Memory-hierarchy highlights}

\textbf{$\\$Cache Organization and management}

Unlike software-managed buffers, caches do not offer the programmer direct control over their content and, specifically, over the choice of item for eviction. Furthermore, in order to simplify their operation and management, caches often restrict the locations in which an item may reside based on certain bits of its original address (the index bits). The number of cache locations at which an item with a given address may reside is referred to as the level of associativity: in a fully associative cache there are no restrictions; at the other extreme, a direct-mapped cache permits any given address to be mapped only to a single specific location in the cache. The collection of cache locations to which a given address may be mapped is called a set, and the size of the set equals the degree of associativity. 

Whenever an item must be evicted from the cache in order to make room for a new one, the cache management system must select an item from among the members of the relevant set. One prominent replacement policy is least recently used (LRU), whereby the evicted item is the set member that was last accessed in the most distant pass. Another is first in -- first out (FIFO), whereby the evicted item is the one that was last brought into the cache in the most distant past. Additional considerations may include eviction of pages that have not been modified while in the cache, because they often don't have to be copied to the lower level in the hierarchy, as it maintains a copy (an inclusive cache hierarchy).

Cache content is managed in units of cache line. We will initially assume that the size of an array item is exactly one cache line, but will later relax this assumption.

\textbf{$\\$Cache performance}

The main cache-performance measure is the hit rate, namely the fraction of accesses that find the desired data in the cache. (Similarly, miss rate = 1- hit rate.)

There are three types of cache misses: Compulsory, Capacity, and Contention \cite{Hennessy}.

1) Compulsory -- a miss that occurs upon the first request for a given data item. (Whenever multiple items fit in a cache line, as well as when automatic prefetching is used, the compulsory miss rate may be lower than expected. Specifically, access to contiguous data would result in one miss per cache line or none at all, respectively.) 

2) Capacity -- this refers to cache misses that would have been prevented with a larger cache. 

3) Conflicts -- these misses occur despite sufficient cache capacity, due to limited flexibility in data placement (limited associativity and non-uniform use of different sets).

\textbf{Remark:} even if cache misses can be reduced by appropriate policies, one must also consider the total communication bandwidth between the cache and the lower level. Specifically, data prefetching can mask latency but does not reduce the mean bandwidth requirement, and speculative prefetching may actually increase it.

\textbf{$\\$Cache coherence}

In multi-core shared memory systems with private caches, yet another complication arises from the fact that the same data may reside in multiple private caches (for reading purposes), yet coherence must be ensured when writing. There are hardware cache-coherence mechanisms that obviate the programmer's need to be concerned with correctness; however, the frequent invocation of these mechanisms can easily become the performance bottleneck. The most expensive coherence-related operations occur when multiple processors attempt to write to the same place. The fact that management and coherence mechanisms operate at cache-line granularity complicates matters, as coherence-related operations may take place even when cores access different addresses, simply because they are in the same cache line. This is known as false sharing.

\textbf{$\\$Cache replacement policy}

A problem may arise at replenishment time. Consider, for example, LRU and a situation wherein a given merge segment only comprises elements of A. As replenishment elements are brought in to replace the used elements of A, the least recently used elements are actually those of B, as both the A element positions and the result element positions were accessed in the previous iteration whereas only one element of B was accessed (it repeatedly ``lost'' in the comparison). A similar problem occurs with a FIFO policy.

A possible solution for LRU is, prior to fetching replenishment elements, touching all cache lines containing unused input elements. If each cache line only contains a single item, this would represent approximately a 50\% overhead in cache access (the usual comparison is between the loser of the previous comparison, which is in a register, and the next element of the winning array, which must be read from cache; also, the result must be written. So the number of cache accesses per step grows from 2 to 3.) If there are multiple elements per cache line, the overhead quickly becomes negligible.

\textbf{$\\$Limited associativity}

\textbf{Proposition} 15. With 3-way associativity or higher, conflict misses can be avoided.

\textit{Proof}: Consider a cache of size $C$. With k-way associativity, any $C/k$ consecutive addresses are mapped to $C/k$ different sets. We partition the merge path into segments of size $C/3$, thus constructing the merged array in segments of $C/3$ elements. The corresponding result array comprises at most $C/3$ elements of A and at most $C/3$ elements of $B$. (The actual numbers are data dependent.) The $C/3$ items of each of $A,B$ and the merged array will take up exactly one position in each of the three sets, regardless of the start address of each of these element sequences. Similarly, each will take up to two positions in a 6-way set associative cache, three in a 9-way, etc.  For associativity levels that are greater than 3 but not integer multiples thereof, one can reduce the segment length such that each array's elements occupy at most a safe number of positions in each set. (A safe number is one such that even if all three arrays occupy the maximum number of positions in a given set, this will not exceed the set size, i.e., the degree of associativity.)  \hfill $\blacksquare$

\begin{figure*}[t]%
\centering
  \includegraphics[width=0.9\columnwidth]{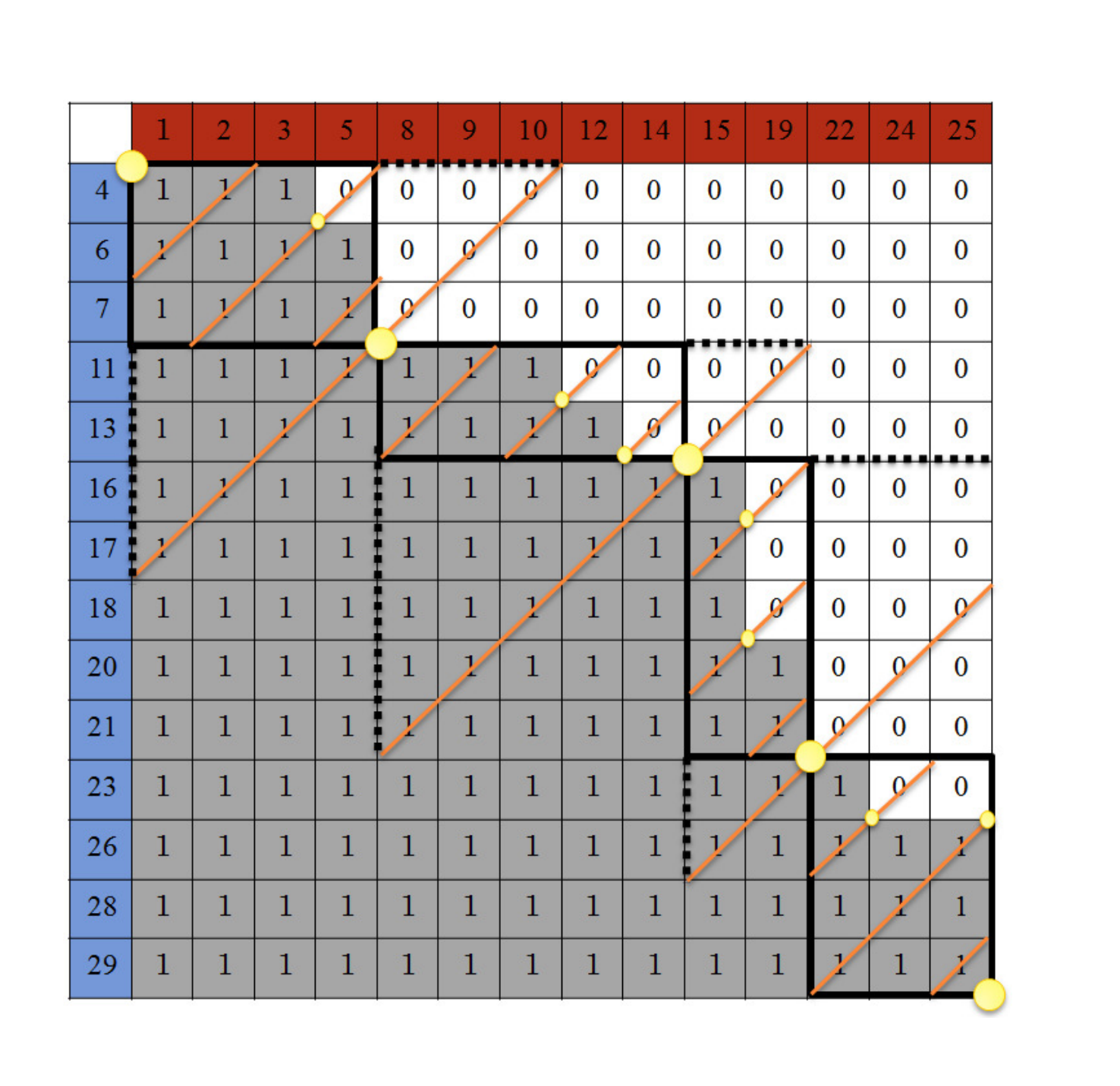}
  \label{fig:CacheMerge}
  \caption{Merge Matrix for the cache efficient algorithm. The yellow circles depict the initial and final points of the path for a specific block in the cache algorithm.}
\end{figure*}

\subsection{ Cache-Efficient Parallel Merge}

In this sub-section we present an extension to our algorithm for parallel merging that is also cache-efficient and considers a shared-memory hierarchy (including a shared cache). 

Collisions in the cache between any two items are avoided when they are guaranteed to be able to reside in different cache locations, as well as when they are guaranteed to be in the cache at different times. In a Merge operation, a cache-resident item is usually required for a very short time, and is used only once. However, many items are brought into the cache. Also, the relative addresses of ``active'' items are data dependent. This is true among elements of different arrays (A, B, S) and, surprisingly, also among same-array elements accessed by different cores. This is because the segment-partition points in any given array are data dependent, as is the rate at which an array's elements are consumed.

Given our efficient parallelization, we are able to efficiently carry out parallel merger of even cache-size arrays. In view of this, we explore approaches that ensure that all elements that may be active at any given time can co-reside in cache. 

Let $C$ denote cache size (in elements). Our general approach is to break the overall merge path into cache-size (actually a fraction of that) segments, merging those segments one after the other, with the merging within each segment being parallelized. We refer to this as Segmented Parallel Merge (SPM). See Fig. 2. The pseudo code for SPM is given in Algorithm \ref{alg:MergePathSegmented}.

\textbf{Lemma 16}. A merge-path segment of length $L$ comprises at most $L$ consecutive elements of $A$ and at most $L$ consecutive elements of $B$.             \hfill $\blacksquare$

\textbf{Theorem 17}. Given $L$ consecutive elements of $A$ and $L$ consecutive elements of $B$, starting with the first element of each of them in the segment being constructed, one can compute in parallel the p segment starting points so as to enable p consecutive segments of length $L/p$ to be constructed in parallel.

Proof:  Consider the $p-1$ cross diagonals of the merge matrix comprising the aforementioned elements of the two arrays, such that the first one is $L/p$ away from the upper left corner and the others are spaced with the same stride. The farthest cross diagonal will require the $L'th$ provided element from each of the two arrays, and no other point along any of the diagonals will require ``later'' elements. Also, since the farthest diagonal is at distance $L$ from the upper left corner (Manhattan distance), the constructed segment will be of length $L$.              \hfill $\blacksquare$

\textbf{Remark}. Unlike the case of a full merger of two sorted arrays of size $L$, not all elements will be used. While $L$ elements will be consumed in the construction of the segment, the mix of elements from $A$ and from $B$ is data dependent. 

\begin{algorithm}[t]
  \caption{$Segmented Parallel Merge(A,B,S,p,C)$}
  \begin{algorithmic}
  \STATE $L \leftarrow C/3$
  \STATE $length \leftarrow L/p$
  \STATE $MAX_{iterations} \leftarrow 3 \cdot (|A|+|B|) / C$
  \STATE $startingPoint \leftarrow $ top left corner

	\FOR {$k=1$ $to$ $MAX_{iterations}$}
    \FOR{$i=1$ $to$ $p$ $\mathbf{parallel}$}
    	\STATE $DiagonalNum \leftarrow (i-1) \cdot (L)/p +1$
    	\STATE $length \leftarrow (L)/p$
			\STATE $a_{i,start},b_{i,start} \leftarrow $ Compute intersection of the Merge path with DiagonalNum using a binary search
    	\STATE $s_{i,start} \leftarrow startingPoint + (i-1) \cdot (L)/p +1$   	
			\STATE $Merge(A,a_{i,start},B,b_{i,start},S,s_{i,start},length)$
    \ENDFOR
    \IF{$k=p$}
    	\STATE update $startingPoint$
    \ENDIF
		\STATE $\mathbf{Barrier}$
	\ENDFOR
		\label{alg:MergePathSegmented}
  \end{algorithmic}			
\end{algorithm}

In order to avoid the extra complexity of using the same space for input elements and for merged data, let $L=C/3$, where C is the cache size.

\textbf{Remark. }Sufficient total cache size does not guarantee collision freedom (conflict misses can occur). However, we have shown that 3-way associativity suffices to guarantee collision freedom. 

\textbf{Computational complexity} Assuming a total merged-array segment size of $L=C/3$ per sequential iteration of the algorithm, there are $3N/C$ such iterations. In each of those, at most $2L=2C/3$ elements of the input arrays ($L$ of each) need to be considered in order to determine the end of the segment and, accordingly, the numbers of elements that should be copied into the cache. Because the sub-segments of this segment are to be created in parallel, each of the p cores must compute its starting points (in $A$ and in $B$) independently. (We must consider $2L$ elements because the end point of the segment, determined by the numbers of elements contributed to it by $A$ and $B$, is unknown.)

The computational complexity of the cache-efficient merge of $N$ elements given a cache of size $C$ and p cores is:${\rm O}\left({{\rm N}}/{{\rm C}}\cdot {\rm p}\cdot {\rm logC+N}\right).$

Normally, $p<<C<<N$, in which case this becomes $O(N)$. In other words, the parallelization overhead is negligible.

The time complexity is ${\rm O}\left({{\rm N}}/{{\rm C}}\cdot \left({\rm logC+C/p}\right)\right).$

Neglecting $logC$ (the parallelization overhead) relative to $C/p$ (the merge itself), this becomes $O(N/p)$, which is optimal. Finally, looking at typical numbers and at the actual algorithms, it is evident that the various constant coefficients are very small, so this is truly an extremely efficient parallel algorithm and the overhead of partitioning into smaller segments is insignificant.

\begin{figure}[t]%
\centering
  \includegraphics[width=0.9\columnwidth]{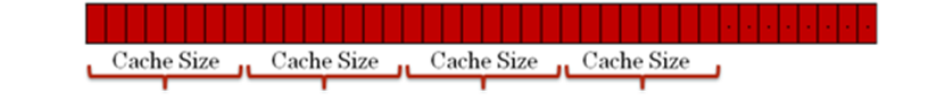}
  \label{fig:Sorting}
  \caption{Cache-efficient parallel sort first stage. Each cache sized block is sorted followed by parallel merging}
\end{figure}

\subsection{ Cache-Efficient Parallel Sort}

Initially, partition the unsorted input array into equisized sub-arrays whose size is some fraction of the cache size $C$.

Next, sort them one by one using the parallel sort algorithm on all p processors as explained in an earlier section. (Of course, one may sort them in parallel, but this would increase the cache footprint.)

Finally, proceed with merge rounds; in each of those, the cache-efficient parallel merge algorithm is applied to every pair of sorted sub-arrays. This is repeated until a single array is produced.  

We now derive the time complexity of the cache efficient parallel sort algorithm. We divide the complexity into two stages: 1) the complexity of the parallel sorting of the sub-arrays of at most $C$elements, and 2) the complexity of the cache-efficient merge stages.

In the first stage, depicted in Fig. 3, the parallel sort algorithm is invoked on the cache sized sub-arrays. The number of those sub-arrays is ${\rm O}\left({\rm N/C}\right)$. Hence, the time complexity of this stage is $O(N/C \cdot (C/p \cdot log(C)+ log(p) \cdot \log  (C)))$.

The second stage may be viewed as a binary tree of merge operations. The tree leaves are the sorted cache sized sub-arrays. Each two merged sub-arrays are connected to the merged sub-array, and so on. The complexity of each level in the tree is $O(log(N/C) \cdot (N/p + N/C \cdot log(p))$.

The total complexity of the cache-efficient parallel sort algorithm is the sum of the complexities of the two stages: 
$ O(N/p \cdot log(N) + N/C \cdot log(p)\cdot log(C))$.

One may observe again that the new algorithm has a slightly higher complexity, ${\rm N/C}\cdot {\log  {\rm (C)}\cdot \ }{{\rm log(} {\rm p)}\ }{\rm >}{\log  {\rm N}\ }\cdot {\log  {\rm (p)}\ }{\rm \ }$, due to the numerous partitioning stages. However, this is beneficial for systems where a cache miss is relatively expensive.

\section{ Related work}

In this section, we review previous works on the subjects of parallel sorting and parallel merging, and relate our work to them. 

Prior works fall into two categories: 1) algorithms that use a problem-size dependent number of processors, and 2) algorithms that use a fixed number of processors. 

Several algorithms have been suggested for parallel sorting. While parallel merge can be a building block for parallel sorting, some parallel sorting algorithms do not require merging. An example is Bitonic Sort \cite{batcher1968sorting} in which ${\rm O}\left({\rm N}\cdot {\left({\log  {\rm N}\ }\right)}^{{\rm 2}}\right)$ comparators are used (${\rm N/2}$ comparators are used in each stage) to sort ${\rm N}$ elements in ${\rm O}\left({\left({\log  {\rm N}\ }\right)}^{{\rm 2}}\right)$ cycles. Bitonic sort falls into the aforementioned first category. Our work is in the latter.

We consider two complexity measures: 1) time complexity (the time required to complete the task), and 2) overall work complexity, i.e, the total number of basic operations carried out. In a load balanced algorithm like ours, the work complexity is the product of time complexity and the number of cores. Even with perfect load balancing, however, one must be careful not to increase the total amount of work (overhead, redundancy, etc.), as this would increase the latency. Similarly, one must be careful not to introduce stalls (e.g., for inter-processor synchronization), as these would also increase the elapsed time even if the ``net'' work complexity is  not increased.

Merging two sorted arrays requires $\Omega {\rm (N)}$ operations. Some of the parallel merging algorithms, including ours, have a work complexity of $O(N+p \cdot log(N))$, the latter component is negligible and the complexity is ${\rm O(N)}$, as observed in \cite{AKL_1987}. Also, there are no synchronization stalls in our algorithm.

In Shiloach and Vishkin \cite{SHIL_VISH}, as in our work, a ${\rm CREW\ PRAM}$ memory model is used. There, a mechanism for partitioning the workload is presented. This mechanism is less efficient than ours and does not feature perfect load balancing; although each processor is responsible for merging ${\rm O}\left({\rm N/p}\right)$ elements on average, a processor may be assigned as many as ${\rm 2N/p}$ elements. This can introduce a stall to some of the cores since all the cores have to wait for the heaviest job. For truly efficient algorithms, namely ones in which the constants are also tight, as is the case with our algorithm, such a load imbalance can cause a 2X increase in latency! The time complexity of this algorithm is ${\rm O}\left({\rm 1+}{\log  {\rm p}\ }{\rm +}{\log  {\rm N}\ }{\rm +N/p}\right)$. For ${\rm N}\gg {\rm p}$, which is the case of interest, it is ${\rm O}\left({\rm N/p+}{\log  {\rm N}\ }\right)$ . 

In \cite{AKL_1987}, Akl and Santoro present a merging algorithm that is memory-conflict free using the ${\rm EREW}$ model. It begins by finding one element in each of the given sorted arrays such that one of those two elements is the median (mid-point) in the output array. The elements found ${\rm (A}\left[{\rm i}\right]{\rm ,B}\left[{\rm j}\right]{\rm )}$ are such that if ${\rm A}\left[{\rm i}\right]$ is the aforementioned median then ${\rm B}\left[{\rm j}\right]{\rm \ }$is the largest element of B that is smaller than ${\rm A}\left[{\rm i}\right]$ or the smallest element of B that is greater than ${\rm A}\left[{\rm i}\right]$. Once this median point has been found, it is possible to repeat this on both sets of the sub-arrays. Their way of finding the median is similar to the process that we use yet the way the explain their approach is different. The complexity of finding the median is ${\rm O(}{\log  \left({\rm N}\right)\ }{\rm )}$. As these arrays are non-overlapping, there will not be any more conflict on accessed data. This stage is repeated until there are ${\rm p}$ partitions. This requires ${\rm O(}{\log  \left({\rm p}\right)\ }{\rm )}$ iterations. Once all the partitions have been found, it is possible to merge each pair of sub-arrays sequentially, concurrently for all pairs, and to simply concatenate the results to form the merged array. The overall complexity of this algorithm is ${\rm O}\left({\rm N/p+}{\log  {\rm N}\ }{\log  {\rm p}\ }\right)$. The somewhat higher complexity is the price for the total elimination of memory conflicts.

In \cite{Deo1991}, Deo and Sarkar present an algorithm that is conceptually similar to that of \cite{AKL_1987} is presented. They initially present an algorithm that finds one element in each of two given sorted arrays such that one of these elements is ${\rm k-th}$ smallest element in the output (merged) array. In \cite{AKL_1987} they start off by finding ${\rm k=N/2}$. In \cite{Deo1991}, the elements sought after are those that are equispaced (${\rm N/p}$ positions apart) in the output array. Finding each of these elements has the complexity of ${\rm O}\left({\log  \left({\rm N}\right)\ }\right)$. This algorithm is aimed for ${\rm CREW}$ systems. The complexity of this algorithm is ${\rm \ O}\left({\rm N/p+}{\log  {\rm (N)}\ }\right)$. 

Our algorithm is very similar to the one presented in \cite{Deo1991}. However, our approach is different in that we show a correspondence between finding the desired elements and finding special points on a grid. Finally, using this correspondence along with additional insights and ideas, we also provide cache efficient algorithms for parallel merging and sorting that did not appear in any of the related works. 

The work done in \cite{DEO_EREW} is an extension of \cite{Deo1991}, in which the algorithm is adapted to an ${\rm EREW}$ machine with a slightly larger complexity of $O(N/P+log(N)+log(P))$ due to the additional constraints of the ${\rm EREW}$. 

In Table \ref{tab:cachemisses}, a summary of the number of cache misses for the aforementioned parallel merge algorithms is given (assuming 3-way associativity as discussed in Proposition 15). Note that the Segmented Merge Path algorithm has a different asymptotic boundary than the remaining algorithms.
This is attributed to the possible sharing of cache lines by different cores that the segmented algorithm does not have, thus Segmented Merge Path has the lowest bound on the number of cache misses in the merging stage. Also note that for Segmented Merge Path there is an overlap of cache misses between the partitioning stage and the merge stage: elements fetched in the partitioning stage will not be fetched again in the merging stage. We remind the reader that $C$ denotes the cache size and the assumption that $p<<C<<N$.

Merging and sorting using GPUs is a topic of great interest as well, and raises additional challenges that need to be addressed. NVIDIA's CUDA  (Compute Unified Device Architecture) \cite{CUDAwebsite} has made the GPU a popular platform for parallel application development. 

In Green et al. \cite{GreenMcColl}, the first GPU parallel merge algorithm is presented based on the Merge Path properties. The GPU algorithm uses the Merge Path properties at two different level of granularity. 
The reader is referred to \cite{nvidiaprogramming} for an extended discussion on the GPU; for the purpose of this discussion we simplify the GPU's programming model to two levels, the stream multiprocessors(SM) and stream processors (SP). The SMs resemble the multiprocessors of the x86. The SPs are light-weight thread computing units grouped together with a single instruction decoder. 
  
Green \emph{et al.} use the cross diagonal search for partitioning the work to the GPU's multi processors similar to the process that is done by our algorithm. In the second phase, the SP's of each SM repeat the cross diagonal intersection with a smaller subset of the path using an algorithm that is similar to the cache-efficient algorithm that is presented in this paper such that in each iteration a subset of $A$ and $B$ are fetched by the SPs in a way that the best utilizes the GPU's memory control unit. Once the segments have been fetched into the memory the cross diagonal intersection is repeated for all the SPs on the SM. At time of publication, this is the fastest known algorithm for merging on the GPU.

In \cite{GHS2008} a radix sort for the GPU is presented. In addition to the radix sort, the authors suggest a merge-sort algorithm for the GPU, in which the a pair-wise merge tree is required in the final stages. In \cite{Sintorn20081381}, a hybrid sorting algorithm is presented for the GPU. Initially the data is sorted using bucket sort and this is followed by a merge sort. The bucket approach suffers from workload imbalance and requires atomic instructions (i.e., synchronization). 

Another focus of sorting algorithms is finding a way to implement them in a cache oblivious \cite{Vitter1998} way. As the algorithm in this paper focused on the merging stage and not the entire sort and presented a cache aware merging algorithm, we will not elaborate on cache oblivious algorithms. The interested reader is referred to \cite{chowdhury2010oblivious,cole2010resource,Frigo1999}.

\begin{table}
\caption{Cache misses for parallel merging algorithms assuming a 3-way associativity.}
\label{tab:cachemisses}
\begin{tabular}{|p{1.2in}|p{1.4in}|p{0.7in}|p{1.4in}|} \hline 

 & \multicolumn{3}{|p{3.0in}|}{\textbf{Cache misses }} \\ \hline 
\textbf{Algorithm} & \textbf{\textit{Partitioning stage}} & \textbf{\textit{Merge stage}} & \textbf{\textit{Total}} \\ \hline 

\cite{SHIL_VISH}  & $O(p \cdot log(N) + p \cdot log(p))$ & $ \Omega(N)$ & $O(N + p \cdot log(N) + p \cdot log(p))$  \\ \hline 
\cite{AKL_1987}  & $O(p \cdot log(N))$ & $ \Omega(N)$ & $O(N + p \cdot log(N))$  \\ \hline 
\cite{Deo1991} \& Merge Path  & $O(p \cdot log(N))$ & $ \Omega(N)$ & $O(N + p \cdot log(N))$  \\ \hline 
Segmented Merge Path  & $O(p \cdot N/C \cdot log(C))$ & $ \Theta(N) $ & $\Theta(N)$  \\ \hline 

\end{tabular}
\end{table}

\section{Results}

\begin{figure}[t]
	\centering
		\includegraphics{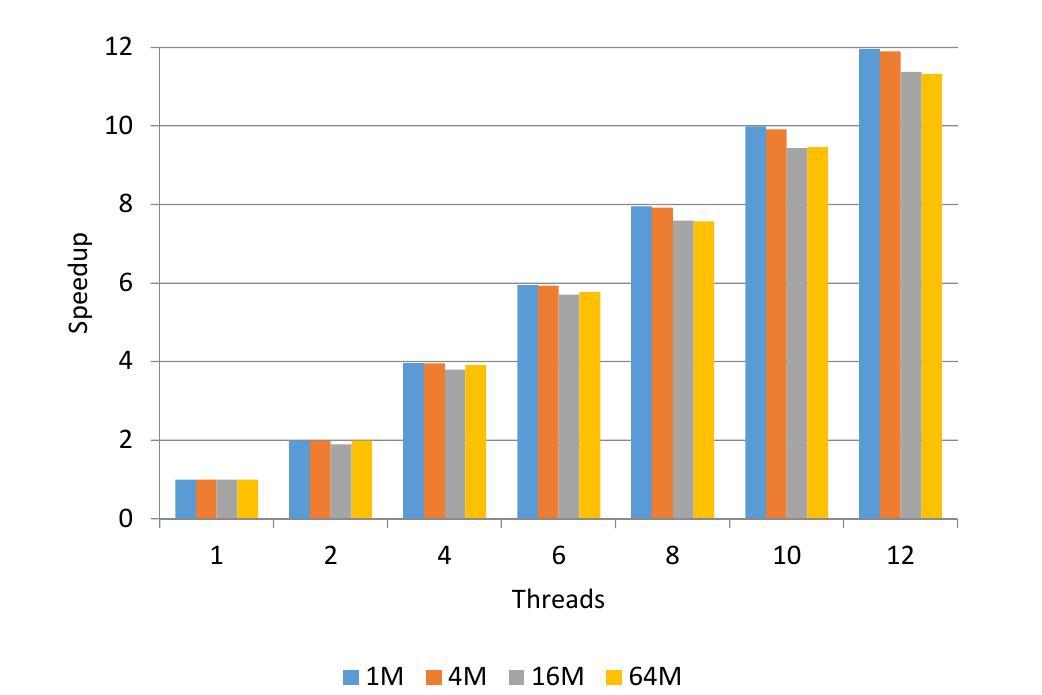}
	\label{fig:dell_12_core}
	\caption{Merge Path speedup on a 12-core system. These results can be found in \cite{MergePath}. The different color bars are for different sized input arrays. 1M elements refers to $2^{20}$ elements. For each array size, the arrays are equisized.}
\end{figure}

According to Amdahl's Law \cite{amdahl1967validity}, a fraction of an algorithm that is not parallelized limits the possible speedup, It is quite evident from the previous sections that we have succeeded in truly parallelizing the entire merging and sorting process, with negligible overhead for any numbers of  interest. Nonetheless, we wanted to obtain actual performance results on real systems, mostly in order to find out whether there are additional issues that limit performance. Also, so doing increases the confidence in the theoretical claims.

We implemented our basic Merge Path algorithm and the cache-efficient version. In the latter, the two arrays are segmented as described such that each segment-pair fits into a 3-way associative cache with no collisions, and the merging of one such segment pair begins only the merging of the previous pair has been completed. However, the actual fetching of array elements into the cache is done only once demanded by the processor, though any prefetch mechanisms of the system may kick in.

The algorithms were implemented on two very different platforms: the x86 platform and the HyperCore, a novel shared-cache many-core architecture by Plurality. We begin with a brief overview of the two systems systems, including system specifications, and then present some of the practical challenges of implementing the algorithms on each of the platforms. Following this, we present the speedup of both the new algorithms, regular Merge Path and the cache efficient version, on each of the systems. The runtime of Merge-Path with a single thread is used as the baseline.

\begin{table}[t]
\caption{Intel X86 systems used}
\label{tab:processors}
\scriptsize	
\begin{tabular}{|c|c|c|c|c|c|c|c|} \hline 

Proc. & $\#$Proc & $\#$ Cores Per Proc. & Total Cores & L1 & L2 & L3 & Memory  \\ \hline

X5670 Intel & 2 & 6 & 12 & 32KB & 256KB & 12MB & 12GB \\ \hline
E7-8870 Intel & 4 & 10 & 40 & 32KB & 256KB & 30MB & 256GB \\ \hline

\end{tabular}
\end{table}

\begin{figure}[t]%
  \centering
	\subfloat[$|A|=|B|=10M$ with full writes to \newline the output array using NUMA Control.]{\includegraphics[width=0.5\textwidth]{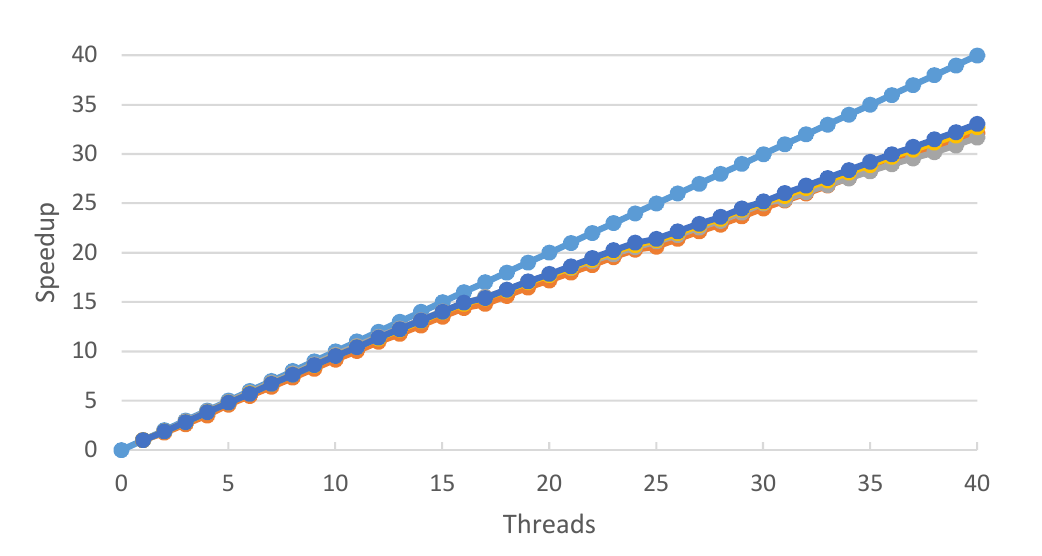}} 
	\subfloat[$|A|=|B|=50M$ with full writes to \newline the output array using NUMA Control.]{\includegraphics[width=0.5\textwidth]{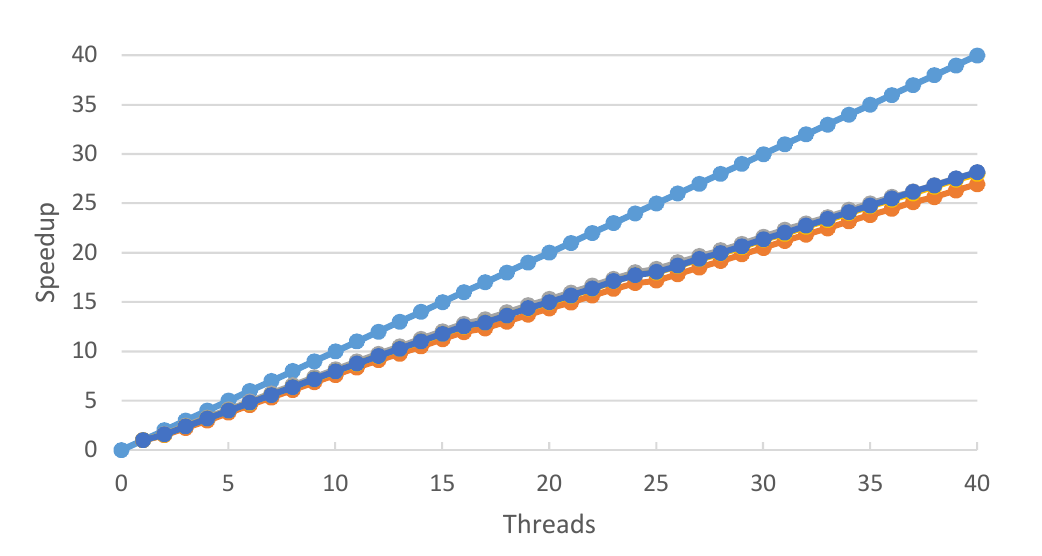}} 

	\subfloat[$|A|=|B|=10M$ with writes to local \newline register.]{\includegraphics[width=0.5\textwidth]{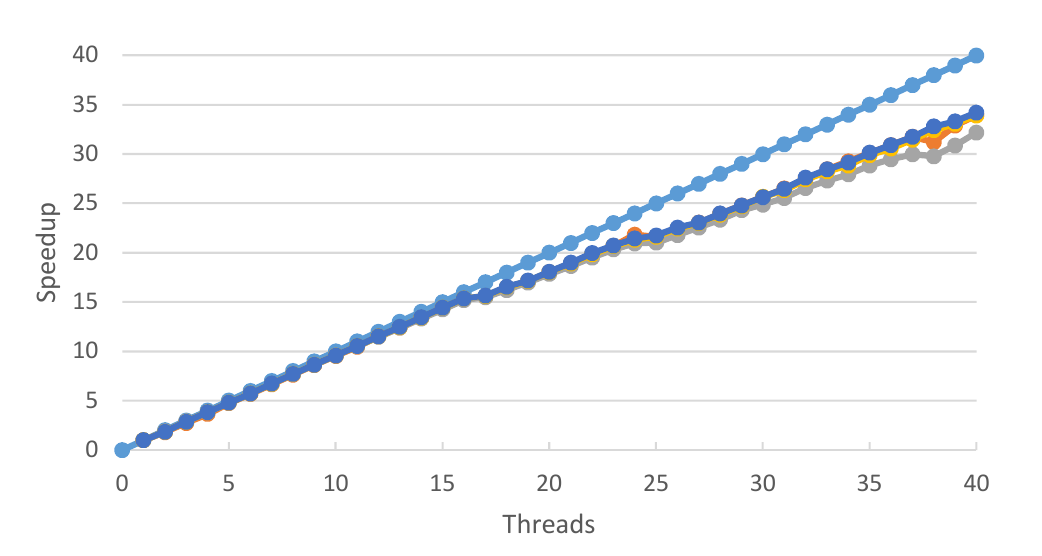}} 
	\subfloat[$|A|=|B|=50M$ with writes to local \newline register.]{\includegraphics[width=0.5\textwidth]{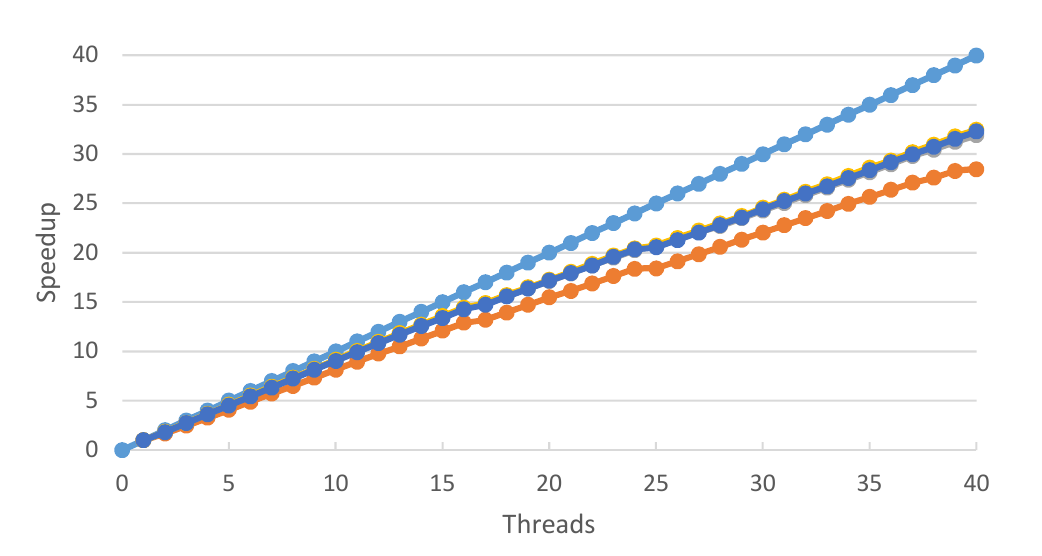}} 

%	\subfloat[$|A|=|B|=10M$ with full writes to \newline the output array.]{\includegraphics[width=0.5\textwidth]{graphics/mir_10_writes.pdf}} 
% 	\subfloat[$|A|=|B|=50M$ with full writes to \newline the output array.]{\includegraphics[width=0.5\textwidth]{graphics/mir_50_writes.pdf}} 
%	\subfloat[$|A|=|B|=10M$ with writes to local register.]{\includegraphics[width=0.5\textwidth]{graphics/mir_10_no_writes.pdf}}

%	\subfloat[$|A|=|B|=10M$ with writes to local register.]{\includegraphics[width=0.5\textwidth]{graphics/mir_10_no_writes.pdf}}

%	\subfloat[$|A|=|B|=50M$ with full writes to \newline the output array.]{\includegraphics[width=0.5\textwidth]{graphics/mir_10_writes.pdf}} 
%  \subfloat[$|A|=|B|=50M$ with writes to local register.]{\includegraphics[width=0.5\textwidth]{graphics/mir_10_no_writes.pdf}}

	\subfloat{\includegraphics[width=\textwidth]{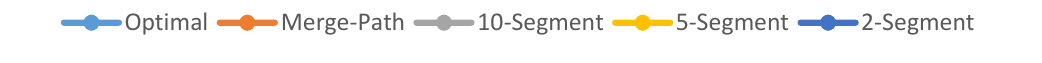}} 

	\caption{Merge Path and Segmented Merge Path speedup on a 40-core system. 1M elements refers to $2^{20}$ elements. For each array size, the arrays are equisized. For the segmented algorithm, the Merge Matrix is divided into 10, 5, and 2 segments.}
  \label{fig:mirasol}

\end{figure}

\subsection{x86 System}

We used two different Intel X86 with Hyperthreading support: a 12 core system and a 40 core system. For both systems, each core has a L1 and L2 private cache. The cores on each processor share a L3 cache. The specifics of these systems can be found in Table \ref{tab:processors}. 
Because the cores have private caches, a cache coherency mechanism is required to ensure correctness.  Furthermore, as we had multiple processors, each with its own L3 cache, the cache coherence mechanism had to communicate across processors; this is even more expensive from a latency point of view. We do not use the Hyperthreading. For each array size and each thread count, multiple executions were completed and the average was taken.

Our implementation of Merge Path is OpenMP \cite{dagum1998openmp} based. On the 12-core system we tested the regular algorithm. On the 40-core system we tested both algorithms (regular and segmented). For both systems we use multiple sizes of arrays and thread count.

Fig. \ref{fig:dell_12_core} depicts the speedup of the algorithm as a function of threads for the 12-core system. Each of the different colored bars is for a different input size. This figure can also be found in our previous paper, \cite{MergePath}. In all cases $|A|=|B|$.
The output array $S$ is twice this size, meaning that the total memory required for the 3 arrays is $4 \cdot |A| \cdot |type|$, where $|type|$ denotes the number of bytes need to stored the data type (for 32 bit integers this will be 4).
One mega element refers to $2^{20}$ elements.  As can be seen, the speedups are near linear, with a slight reduction in performance for the bigger input arrays: approximately $11.7X$ for $12$ threads. 

%Remark. We note that the single-thread execution time of our algorithm was some $6\%$ longer than a truly sequential merge algorithm. This is due in part to a few extra instructions, and possibly also to overhead of OpenMP.

Fig. \ref{fig:mirasol} depicts the speedup on a larger 40-core system for both the regular and segmented algorithm. For the segmented algorithm we used multiple segment sizes: two, five, and ten segments. As such each segment size is $|S|/\#segments$. 
Effective parallelization on the 40 core system is more challenging than it is for the 12 core system as the 40 core system consists of 4 processors (each with 10 cores). The 4 processor design can potentially add overhead related to synchronization and cache coherency. Additional considerations include memory bandwidth saturation due to the algorithm being communication bound. We discuss these.

For each input size tested, we present two sets of results - execution time with the write backs and execution time without write backs.
Both use the NUMA Contral package \cite{kleen2005numa}.
The speedups for the writeback of the results to the main memory are depicted in Fig. \ref{fig:mirasol}(a) and Fig. \ref{fig:mirasol}(b), for array sizes 10M and 50M respectively. The speedups for the algorithms when the results are written to a register are depicted in Fig. \ref{fig:mirasol}(c) and Fig. \ref{fig:mirasol}(d), for array sizes 10M and 50M respectively. 

Note that for both array sizes, the speedup attained at 10 threads is not doubled at 20 threads and the speedup at 20 threads is not  doubled for 40 threads. There are two causes for this: writing back of the elements and synchronization.

As expected, writing the results back to the main memory adds latency and thus reduces the speedup from about 32X to 28X for the 40-thread execution for the $50M$ array size. This side effect is not felt as much for the smaller array sizes, $10M$, a significant part of the array fits in the caches. Note that each processor has a $30MB$ L3 cache, with a total of $120MB$ for all 4 processors. The amount of memory required by the $A,B,$ and $S$ is $160MB$. As such, at the end of the merge, part of the array is still in the cache and is not written back to the main memory as occurs for the larger arrays. This behavior is not a side affect of our algorithm, rather it is a side affect of the architecture.

To verify the impact of the synchronization, we timed only the path intersection and its immediate synchronization (meaning that merging process itself was not timed). As the path intersection takes at most $O(log_2(min(|A|,\\ |B|)))$ steps, if the main cross diagonal is used for intersection in the case of equisized inputs, this intersection should take the longest amount of time. For example, when 2 threads are used, the $2^{nd}$ thread uses the main cross diagonal. Meaning it should be the bottleneck in the computation. As we increased the number of threads, the amount of time to find the intersections grew - regardless of the amount of computation needed. For the segmented algorithm the synchronization time also grew and became more substantial. Once again, this is not a side affect of our algorithm.

We now compare the regular algorithm with the segmented algorithm.
For the smaller array, the segmented algorithm is slightly outperformed by the regular algorithm. This is due to the synchronization overhead where the synchronization costs plays a more substantial part of the total execution time.
Also, likelihood of cache contention is smaller for these arrays. As such the regular algorithm outperforms the segment algorithm. 
As the sizes of the array increases so does the likelihood of contention (for both the partitioning and merging stages). It is not surprising that the new cache efficient algorithm outperforms the regular algorithm.

In summary, our x86 OpenMP implementations perform rather well and achieves $75\%-90\%$ of the full system utilization (problem size dependent).

\subsection{HyperCore Architecture}

\begin{figure}
	\centering
		\includegraphics[width=\textwidth]{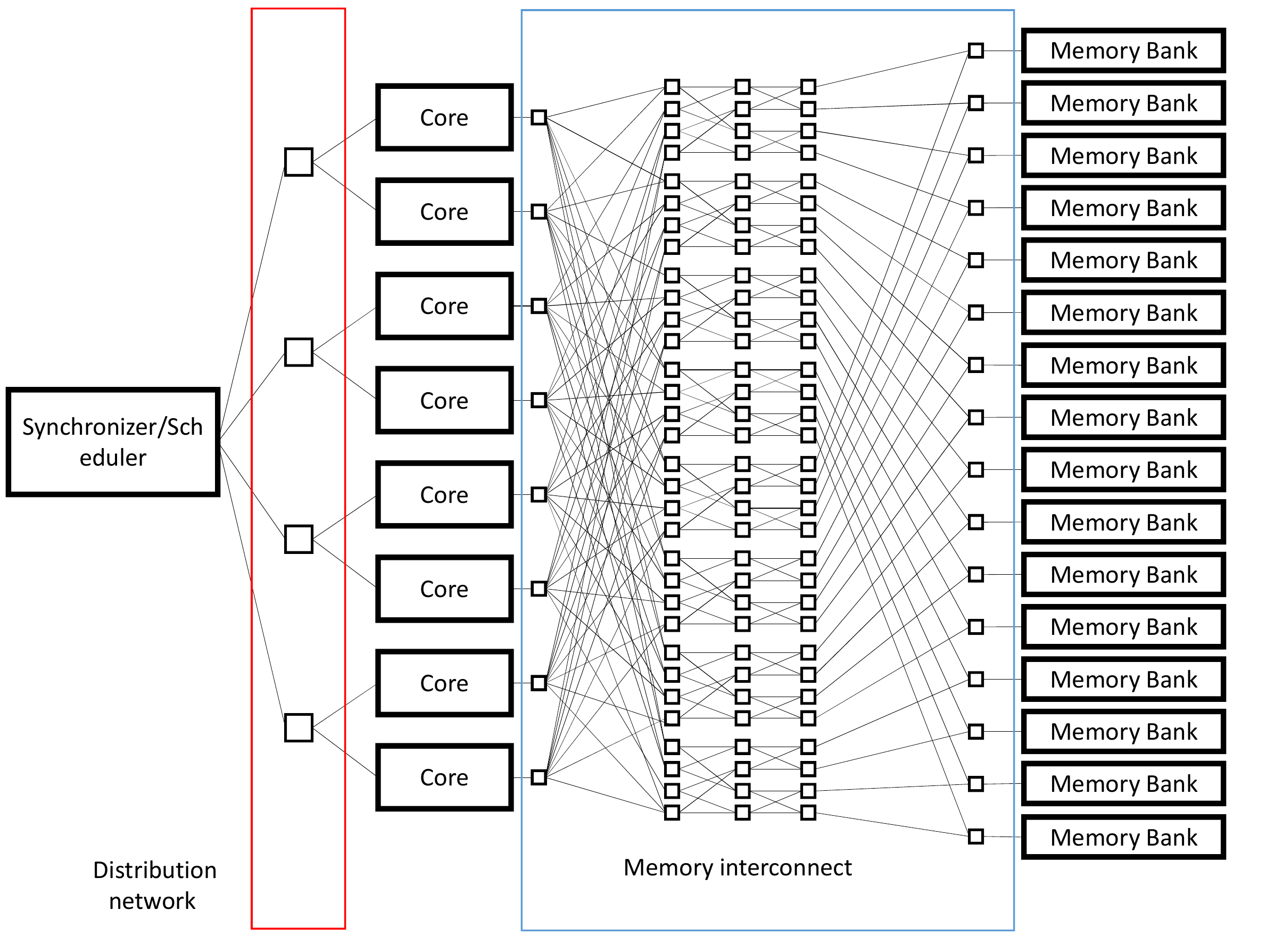}
	\caption{Schematic view of the Hypercore for an 8-core system. The memory banks refer to the shared memory. The Hypercore also has DRAM memory which is not shown in this schematic.}
	\label{fig:hypercore}
\end{figure}

\begin{figure}%
  \centering
	\subfloat[Regular Algorithm]{\includegraphics[width=0.5\textwidth]{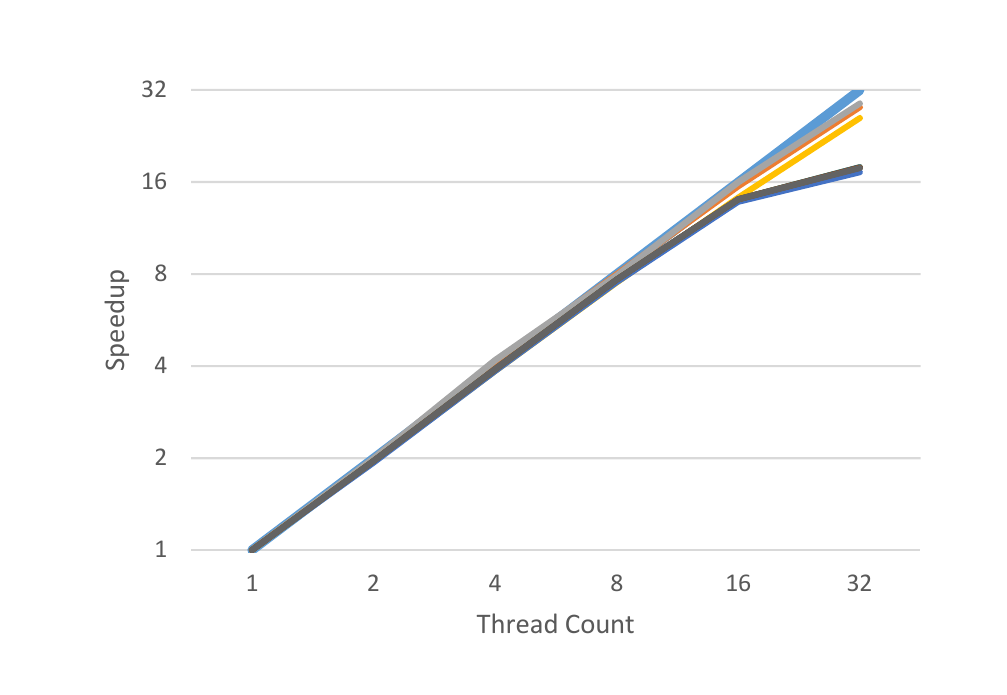}} 
  \subfloat[Segmented Algorithm]{\includegraphics[width=0.5\textwidth]{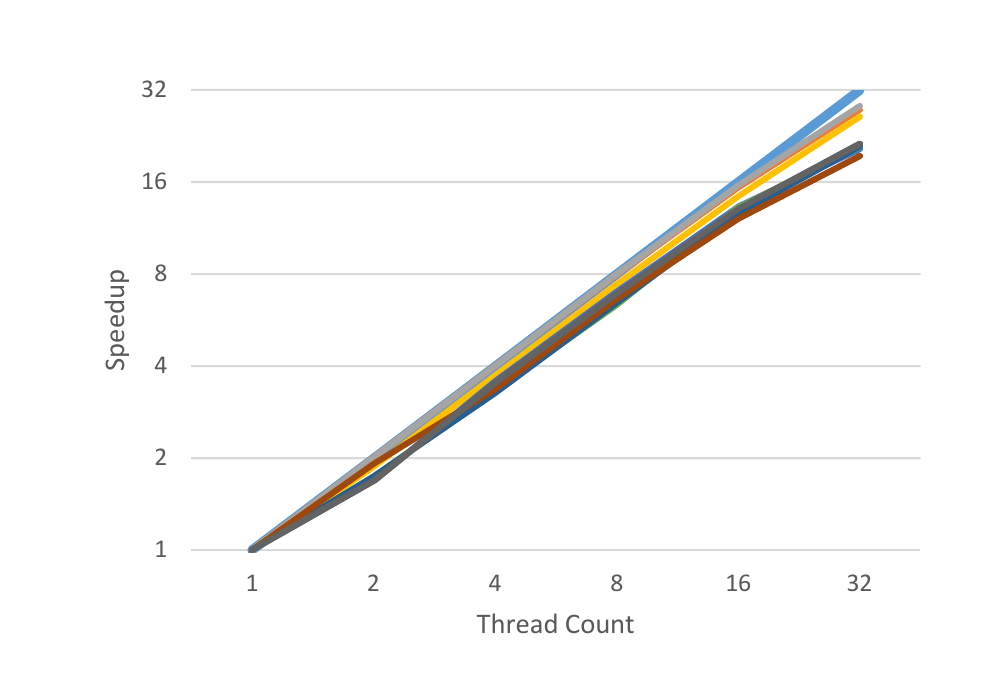}}

  \includegraphics[width=0.8\textwidth]{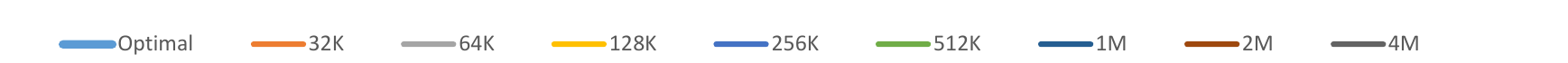}

  \caption{Speedup of our algorithm on Plurality's Hypercore \cite{PluralityEETimes} system.}
  \label{fig:plurality}

\end{figure}

\begin{figure}%
  \centering

  \includegraphics[width=0.5\textwidth]{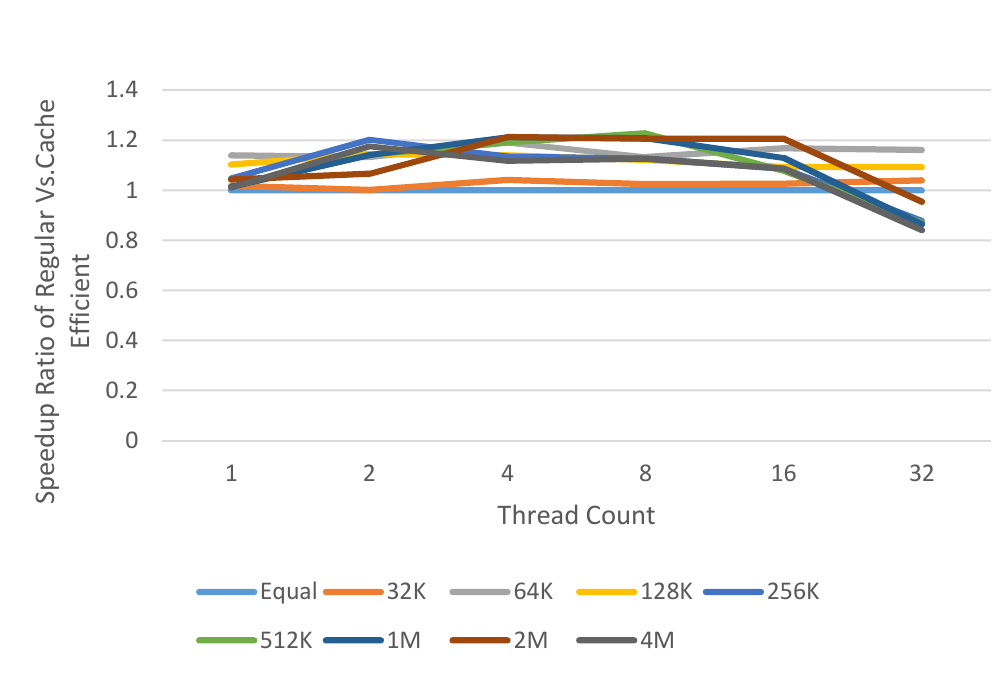}

  \caption{Speedup comparison of the regular and segmented algorithms on the Hypercore. A blue curve, \emph{Equal}, has been added to this chart to denote when the speedups of the algorithm are equal. All curves above \emph{Equal} mean that the regular algorithm outperforms that segmented algorithm. All curves below \emph{Equal} mean that the segmented algorithm is more efficient. }
  \label{fig:plu_ratio}
\end{figure}

Plurality's HyperCore architecture \cite{PluralityHandbook,PluralityEETimes} features tens to hundreds of compute cores, interconnected to an even larger number of memory banks that jointly comprise the shared cache. The connection is via a high speed, low latency combinational interconnect. As there are no private caches for the cores, memory coherence is not an issue for CREW like algorithms. Same-address writes are serialized by the communication interconnect; however, for our algorithm this was not needed. The memory banks are equidistant from all the cores, so this is a UMA system. The shared cache has a number of memory banks that is larger than the number of cores in the system, reducing the number of conflicts on a single bank. Moreover, addresses are interleaved, so there are no persistent hot spots due to regular access patterns. Fig. \ref{fig:hypercore} depicts an 8-core Hypercore system.

The benefit of such an architecture is that there is no processor-cache communication bottleneck. Finally, the absence of private caches (and a large amount of state in them) and the UMA architecture permit any core to execute any compute task with equal efficiency. The memory hierarchy also includes off-chip (shared) memory. Finally, the programming model is a set of sequential "tasks" along with a set of precedence relations among them, and these are enforced by a very high throughput, low latency synchronizer/scheduler that dispatches work to the cores. 

At the time of submission, Plurality has not manufactured the actual chip. We had access to an advanced experimental version of the HyperCore on an FPGA card.  The FPGA version we used has a 1MB direct mapped cache and 32 cores. Furthermore, the FPGA has a latency issue on memory write back. Therefore, results are shown for an algorithm that does not  write to memory. Instead, we saved the value in a private register. 

We ran both the non-segmented and segmented versions of Parallel Merge Path with varying numbers of threads (cores). The input arrays (of type integer) tested on Plurality are substantially smaller than those that we tested on the x86-system due to the FPGA limitations. One might expect that merging smaller arrays would not offer significant speedups due to the overhead required in dispatching threads and to the fact that the search for partition points (binary search on a cross diagonal) become a more significant part of the computation. However, due to HyperCore's ability to dispatch a thread within a handful of cycles, the overhead is not a problem and makes the HyperCore an idle target platform. The sizes of the input arrays are denoted by the number of elements in each of the arrays $A$ and $B$. 
Again, the arrays $A$ and $B$ are equisized. 

Fig. \ref{fig:plurality}(a) presents the speedup of our basic Parallel Merge algorithm as a function of the number of cores. Multiple input sizes were tested. It is evident that they speedup is quite close to linear up to 16 cores, regardless of the array sizes.
For the larger input arrays, the speedup does decrease for the 32 core count. This is most likely due to shared-memory contention and does not occur for the segmented algorithm.

Fig. \ref{fig:plurality}(b) similarly depicts the speedup for the segmented algorithm. Note, however that the cache is direct mapped, so collision freedom cannot be guaranteed. Nonetheless, the partition into sequential iteration, each carrying out parallel merge on a segment, does improve performance. The percentage speedup of the segmented version relative to same-parameter execution without segmentation is depicted in Fig. \ref{fig:plu_ratio}. 

%\textbf{Remark.} Note that this is not directly related so speedup over a single core, as even single-core performance is affected. (In other words, Fig. \ref{7} is independent of Fig. 5, 6.)

\section{ Conclusions}

In this paper, we explored the issue of parallel sorting through the cornerstone of many sorting algorithms -- the merging of two sorted arrays.

One important contribution of this paper is a very intuitive, simple and efficient approach to correctly partitioning each of two input sorted arrays into segments that, once pairs of segments, one from each, are merged, the concatenation of the merged pairs yields a single sorted array. This partitioning is also done in parallel.

Another important contribution is an insightful consideration of cache related issues. These are extremely important because, especially when parallelized, sorting and merging are carried out at a speed that is very often determined by the memory subsystem rather than by the compute power.  This culminated in a cache-efficient parallel merge algorithm. To this end, the efficient segmented version of our algorithm is very promising, as it can operate efficiently with simple caches.

%% The Appendices part is started with the command \appendix;
%% appendix sections are then done as normal sections
%% \appendix

%% \section{}
%% \label{}

%% References
%%
%% Following citation commands can be used in the body text:
%% Usage of \cite is as follows:
%%   \cite{key}          ==>>  [#]
%%   \cite[chap. 2]{key} ==>>  [#, chap. 2]
%%   \citet{key}         ==>>  Author [#]

%% References with bibTeX database:

\bibliographystyle{elsarticle-num}
\bibliography{MergePath}

%% Authors are advised to submit their bibtex database files. They are
%% requested to list a bibtex style file in the manuscript if they do
%% not want to use model1a-num-names.bst.

%% References without bibTeX database:

% \begin{thebibliography}{00}

%% \bibitem must have the following form:
%%   \bibitem{key}...
%%

% \bibitem{}

% \end{thebibliography}

\end{document}